\documentclass[conference]{IEEEtran}

\def\BibTeX{{\rm B\kern-.05em{\sc i\kern-.025em b}\kern-.08em
    T\kern-.1667em\lower.7ex\hbox{E}\kern-.125emX}}

\usepackage{cite}
\usepackage{amsmath,amssymb,amsfonts}
\usepackage{textcomp}
\usepackage{xcolor}
\usepackage{xspace}
\usepackage{url}
\usepackage{hyperref}
\usepackage{tabularx}
\usepackage{listings}
\usepackage{listingsutf8}

\usepackage{tikz}
\usepackage{amsmath}
\usepackage{pifont}

\usepackage{graphbox} %
\usepackage{color}
\usepackage{colortbl}

\usepackage{url}
\usepackage{multicol}
\usepackage{multirow}
\usepackage{subcaption}
\usepackage{graphicx} 
\usepackage{algorithmicx} 
\usepackage{algorithm2e}
\PassOptionsToPackage{hyphens}{url} %
\usepackage{hyperref}
\hypersetup{pdfauthor={Name}}
\hypersetup{colorlinks,linkcolor=[rgb]{.75,0,0},urlcolor=[rgb]{0.75,0,0.75},citecolor=blue,breaklinks} %
\usepackage{tikz} 
\usepackage{pgf}
\usetikzlibrary{arrows,automata}
\usepackage[all]{xy}
\usepackage{pbox} 
\usepackage{comment}
\usepackage{amsfonts}
\usepackage{adjustbox}
\usepackage{listings}
\usepackage{xcolor}
\usepackage{fancyvrb}
\usepackage{color}
\usepackage{listings}
\usepackage{tabularx}
\usepackage{fancyvrb}
\usepackage{listingsutf8}
\usepackage{booktabs}
\usepackage[utf8x]{inputenc}
\usepackage{array}

\usepackage[]{footmisc}

\newcommand\reviewfix[1]{{\sffamily [RF:#1]}\@bsphack\@esphack}

\newcommand{\ok}{\checkmark}
\newcommand{\eg}{{\emph{e.g.}}}
\newcommand{\ie}{{\emph{i.e. }}}
\newcommand{\aka}{{\emph{a.k.a.}}}
\newcommand{\verfploeter}{{\sc Verfploeter}\xspace}
\newcommand{\tangled}{{\sc Tangled}\xspace}
\newcommand{\authoremail}[1]{\texttt{\small{#1}}}

\usepackage{listings}
\usepackage{xcolor}

\colorlet{punct}{black!60!black}
\definecolor{background}{HTML}{EEEEEE}
\definecolor{delim}{RGB}{20,105,176}
\colorlet{numb}{magenta!60!black}

\lstdefinelanguage{json}{
    basicstyle=\small\ttfamily,
    numbers=none,
    numberstyle=\scriptsize,
    stepnumber=1,
    numbersep=8pt,
    showstringspaces=false,
    breaklines=true,
    frame=lines,
    backgroundcolor=\color{background},
    literate=
     *{0}{{{\color{numb}0}}}{1}
      {1}{{{\color{numb}1}}}{1}
      {2}{{{\color{numb}2}}}{1}
      {3}{{{\color{numb}3}}}{1}
      {4}{{{\color{numb}4}}}{1}
      {5}{{{\color{numb}5}}}{1}
      {6}{{{\color{numb}6}}}{1}
      {7}{{{\color{numb}7}}}{1}
      {8}{{{\color{numb}8}}}{1}
      {9}{{{\color{numb}9}}}{1}
      {:}{{{\color{punct}{:}}}}{1}
      {,}{{{\color{punct}{,}}}}{1}
      {\{}{{{\color{delim}{\{}}}}{1}
      {\}}{{{\color{delim}{\}}}}}{1}
      {[}{{{\color{delim}{[}}}}{1}
      {]}{{{\color{delim}{]}}}}{1},
}

\lstdefinelanguage{bshell}{
    basicstyle=\ttfamily\scriptsize,
    numbers=none,
    numberstyle=\ttfamily\scriptsize,
    stepnumber=1,
    numbersep=4pt,
    showstringspaces=false,
    breaklines=true,
    frame=lines,
    backgroundcolor=\color{background},
    literate=
      {:}{{{\color{punct}{:}}}}{1}
      {,}{{{\color{punct}{,}}}}{1}
      {\{}{{{\color{delim}{\{}}}}{1}
      {\}}{{{\color{delim}{\}}}}}{1}
      {[}{{{\color{delim}{[}}}}{1}
      {]}{{{\color{delim}{]}}}}{1},
}

\begin{document}
\title{Tangled: A Cooperative Anycast Testbed}

\author{
    \IEEEauthorblockN{
        Leandro M. Bertholdo\IEEEauthorrefmark{1}, 
        João M. Ceron\IEEEauthorrefmark{2},
        Wouter B. de Vries\IEEEauthorrefmark{3},\\
        Ricardo de O. Schmitt\IEEEauthorrefmark{4},
        Lisandro Zambenedetti Granville\IEEEauthorrefmark{5},
        Roland van Rijswijk-Deij\IEEEauthorrefmark{1},
        Aiko Pras\IEEEauthorrefmark{1}
    }
    \vspace{0.5em}
    \IEEEauthorblockA{\IEEEauthorrefmark{1}
        University of Twente, Enschede, The Netherlands\\
      \authoremail{\{l.m.bertholdo, r.m.vanrijswijk, a.pras\}@utwente.nl}
  }
    \vspace{0.5em}
  \IEEEauthorblockA{\IEEEauthorrefmark{2}
      SIDN Labs, Arnhem, The Netherlands\\
      \authoremail{joao.ceron@sidn.nl}
  }
  \vspace{0.5em}
  \IEEEauthorblockA{\IEEEauthorrefmark{3}
      Tesorion, Enschede, The Netherlands\\
      \authoremail{wouter.devries@tesorion.nl}
  }
  \vspace{0.5em}
  \IEEEauthorblockA{\IEEEauthorrefmark{4}
      University of Passo Fundo, Passo Fundo, Brazil\\
      \authoremail{rschmidt@upf.br}
  }
    \vspace{0.5em}
  \IEEEauthorblockA{\IEEEauthorrefmark{5}
      Federal University of Rio Grande do Sul, Porto Alegre, Brazil\\
      \authoremail{granville@inf.ufrgs.br}
    }
}

\maketitle

\begin{abstract}
Anycast routing is an area of studies that has been attracting
interest of several researchers in recent years. Most anycast studies
conducted in the past relied on coarse measurement data, mainly due to
the lack of infrastructure where it is possible to test and collect
data at same time. In this paper we present Tangled, an anycast test
environment where researchers can run experiments and better
understand the impacts of their proposals on a global infrastructure
connected to the Internet.
\end{abstract}


\section{Introduction}

IP anycast consists in announcing different copies of a service in the
Internet using the same IP address, and trusting the Internet routing
(\eg \ BGP~\cite{rfc4271}) to forward and distribute traffic between
service copies.

Initially proposed in 1993, IP anycast was originally used to help
clients find the best application server in the
Internet~\cite{rfc1546}. Since then, IP anycast has been widely
employed for load
balancing~\cite{katabi2000gia}~\cite{RN220}~\cite{RN273}, in the DNS
infrastructure~\cite{agarwal2001content}~\cite{rfc3258}~\cite{sarat2006use},
and CDN cloud
providers~\cite{RN298}~\cite{RN234}~\cite{RN294}~\cite{RN301},
and, more recently, it has also been studied and deployed for DDoS
mitigation~\cite{Moura16b}~\cite{de2016anycast}~\cite{kuipers17}~\cite{Rizvi19a}~\cite{rizvi2020anycast}.
Today, anycast is used to support hundreds of services across the
Internet~\cite{RN275}~\cite{RN276}.

Although there is a large literature on IP anycast, carrying out
real-world experiments with IP anycast is not an easy task. Typically,
and understandably, operators do not allow for running tests on
production networks and servers; and deploying a meaningfully large
anycast network, consisting of various copies of a service widely and
reasonably distributed across the Internet is beyond reach for most
researchers. Building an IP anycast network is not a technically
challenging task per se (in fact, there are many references and
guidelines on how to do
it~\cite{rfc4786}~\cite{ripe69nat}~\cite{Jafferali2016buildanycast}).
However, the major roadblocks are the cost and time involved in the
process of building a proper anycast network following the same
practices of the industry, and retrieving trusted data from that
network.

Based on experiences of our previous work in IP anycast, we argue that
a testbed deployed in the wild is the most feasible and
technically accurate way to run experiments. Testbeds are usually
built on a collaborative way, where industry and academia together
support research that benefit the Internet operations. Compared to
other approaches and methodologies, such as using third-party datasets
for research, testbeds commonly allow for changes in metrics, which
enables the study of a given subject under different conditions.

In this paper, we introduce
\tangled\footnote{\url{https://www.anycast-testbed.nl}}, a world-wide,
collaborative open-access IP anycast testbed. \tangled ultimately aims
to support research on anycast by academia and industry by making the
deployment of anycast-related experiments viable to the overall
community of network research and operation. Our testbed consists of
various copies (\aka\ anycast instances or anycast sites) distributed
around the globe and co-located under different ASes, as well as a set
of tools to: ({\it i}) provide a programmable anycast traffic
engineering interface, able to control each individual anycast site
visibility; ({\it ii}) map the distribution of traffic from clients to
the anycast sites using million of vantage points; and ({\it iii})
measure and analyze result data from experiments. This paper present
the infrastructure of \tangled as of September 2020. We are constantly
looking for opportunities to expand our testbed by establishing new
partnerships and collaborations, as well as the deployment of new
nodes.

The remainder of this paper is organized as follows. In \autoref{sec:testbed} we describe our testbed technical details on connectivity and infrastructure. In \autoref{sec:tagled-TE} we 
show all preprogrammed testbed traffic engineering features available. 
In \autoref{sec:data} we explain our data collection process and
provided data format. In \autoref{sec:lessons} we state some
experience we learn for running this testbed. In \autoref{sec:related}
we compare \tangled with other infrastructures able to develop 
anycast research.

\section{The Tangled Testbed}
\label{sec:testbed}

Configuring and deploying an anycast network is a process that
involves a constant maintenance. Upstreams and IXPs change policies
and infrastructure from time to time.  However, \tangled active
measurement infrastructure allows to identify BGP routing
configurations mistakes or relevant infrastructure changes made by ISP
or IXPs where we have presence. This capability provide us a more
trustable anycast testbed environment.

\tangled\ consists of thirteen sites, most of these deployed through
partnership with universities and academical networks, registrars, and
transit providers. Some of our anycast sites are deployed within cloud
commercial networks, with the goal to increase the coverage of our
anycast network to regions where we currently have no partners. In the
case of an anycast network for research purposes, we generally believe
that the more sites the better, mainly if these sites are located
within different ASes; more sites in different networks increase, for
example, the possibilities of combinations for experiments and
observation of routing dynamics. Therefore, we believe that
cooperation is a key factor to keep the \tangled testbed growing and
with a meaningful number of relevant sites.

\subsection{Historical Context}

\tangled was conceived in 2016, during a BGP hackathon organized by
CAIDA/UCSD~\cite{dainotti2016hackatoon}. In that event, while
developing their BGP project, the team ``Anycast-1'', with members
from the University of Twente (UT) among others, discovered
misconfigurations within the
\emph{Peering}~\cite{schlinker2014peering} BGP testbed. That situation
helped us understand the challenges on building an anycast network,
and it was the main motivation for the UT researchers to start
planning their own testbed infrastructure. The first release of the
\tangled testbed was publicly presented in 2016, at
RIPE73~\cite{ripe73wouter}.

In the following years, we expanded our community network around the
testbed, deploying anycast sites around the world. Several researches
were carried out along the years using the \tangled network: anycast
catchment studies~\cite{de2019improving} and the tool called
\verfploeter~\cite{devries2017verfploeter}; and several anti-DDoS
studies from~\cite{de2016anycast,kuipers17} were carried out using our
testbed. Moreover, the \tangled testbed is actively being used in the
projects SAND~\cite{sand2020ceron} and
PaaDDoS~\cite{paaddos2020bertholdo}.

\subsection{Addressing Infrastructure}

\tangled has its own AS (1149), and prefixes
(145.100.118.0/23 and 2001:610:900::/40) provided by SURFnet -- the
Dutch NREN. Prefixes are RPKI signed and properly described
on RIRs databases, increasing security of our routing environment and
preventing the prefixes misuse. Multiple distinct experiments can be
configured and executed at the same time in \tangled by using smaller
prefixes; for example announcing two /24 prefixes instead of our
original /23 one, or even a fraction of the IPv6 address space

\subsection{Connectivity}

\begin{figure}
  \includegraphics[width=0.9\linewidth]{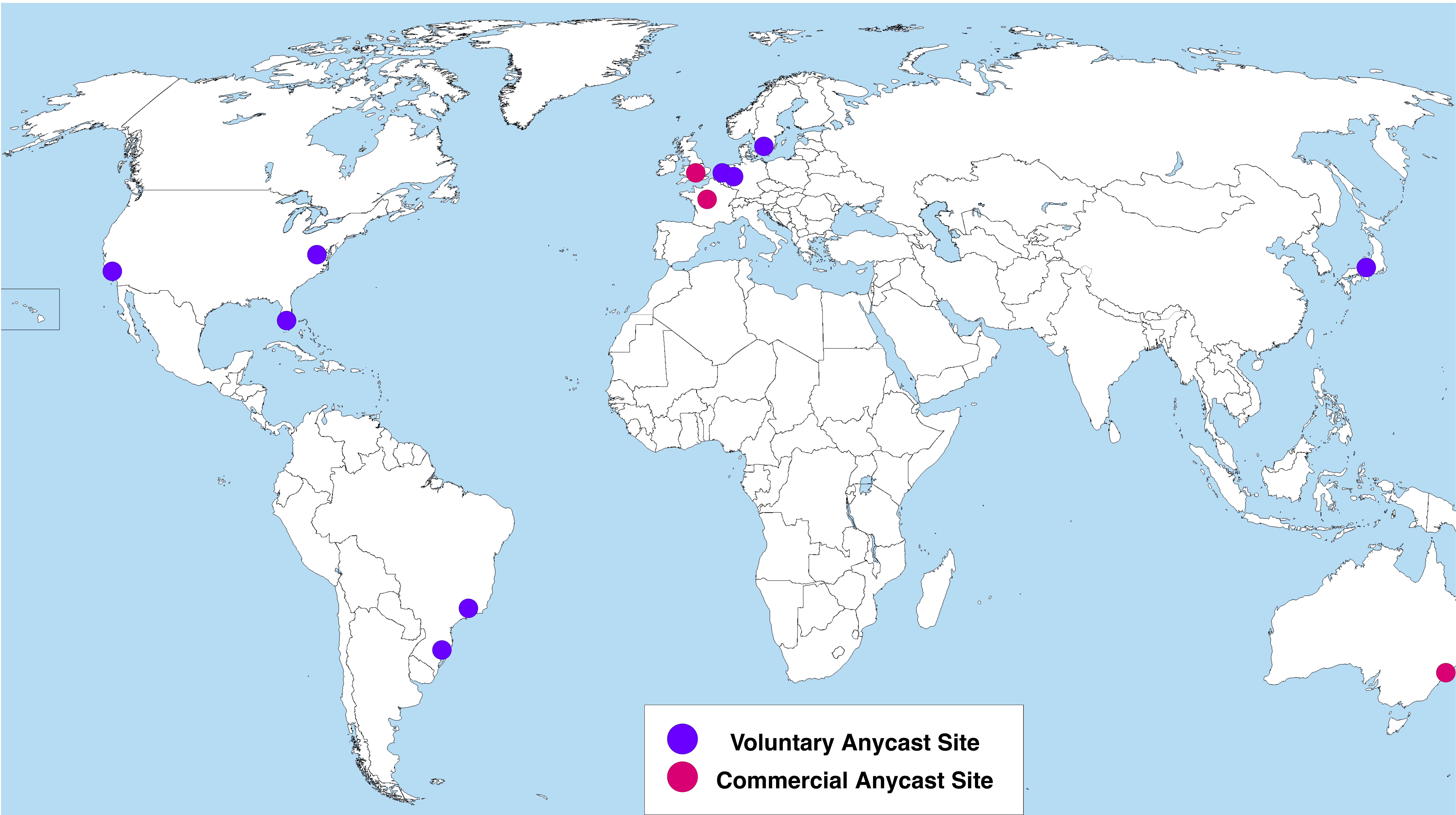}
  \caption{Anycast sites provided by Tangled.}
  \label{fig:tangled-map}
\end{figure}

\tangled has one master site used to consolidate data, and twelve
anycast sites deployed in Asia (1 site), Europe (5), South America
(2), North America (3), and in Oceania (1), as depicted in
\autoref{fig:tangled-map}. Four sites are connected to IXPs, meaning
that these sites have richer connectivity (\ie\ more visibility across
the Internet): both sites in Brazil (São Paulo and Porto Alegre) are
directly connected to the Brazilian Internet Exchange Point (IX.br);
the sites in London and Paris have access to LINX and FranceIX,
respectively. \autoref{tab:tangled-nodes} details our transit
providers and IXP connections. Some of our anycast sites share the
same upstream provider, while others peer with various commercial and
academic networks.

\begin{figure}
 \includegraphics[width=1.00\linewidth]{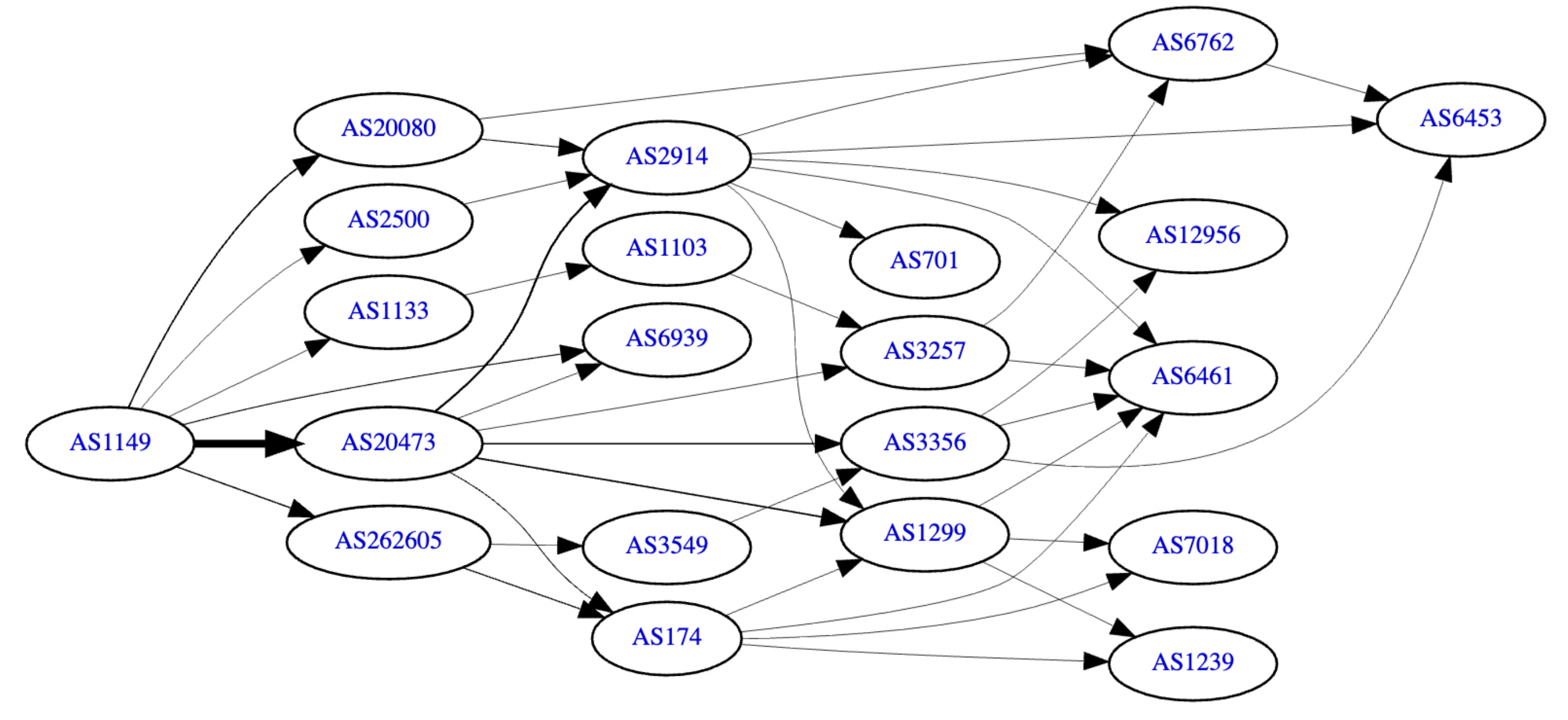}
 \caption{Route propagation map (source:he.net).}
 \label{fig:as1149-map}
\end{figure}

Since site connectivity have a direct relationship with the anycast
catchment\footnote{Anycast catchment is defined by the distribution of
source traffic as defined by BGP routing decisions, ultimately
defining the set of sources an anycast site sees in its incoming
traffic}~\cite{mcquistin2019taming}, \ie\ BGP might prefer to forward
traffic to a more distant site but with better connectivity. This
variety of connectivity provides valuable study cases for the testbed.
\autoref{fig:as1149-map} shows the Huricane Eletric looking glass view
of AS1149.

\begin{table}
  \scalebox{0.9}{%
  \small
  \begin{tabularx}{\textwidth}{lp{18mm}p{23mm}ll}
\cline{1-5}
Site ID   & Location                     & Transit Provider          &  IXP  & Peers  \\ \cline{1-5}
\rowcolor{cyan!10} 
au-syd    & Sidney\newline Australia    & Vultr (20473)             & --   & 1      \\
br-gru    & São Paulo\newline Brazil    & Ampath(20080) ANSP(1251) & spo.IX.br & 1892     \\
\rowcolor{cyan!10}
br-poa    & Porto Alegre\newline Brazil & Leovin(262605) Nexfibra(264575)& poa.IX.br  & 218      \\
dk-cop    & Copenhagen\newline Denmark  & DK-Hostmaster (39839)     & --   & 1      \\
\rowcolor{cyan!10}
uk-lnd    & London\newline England      & Vultr (20473)             & Linx   & 1      \\
fr-par    & Paris\newline France        & Vultr (20473)             & France-IX   & 1      \\
\rowcolor{cyan!10}
jp-hnd    & Tokyo\newline Japan             & Wide (2500)           & --   & 1      \\
nl-ens    & Enschede\newline Netherlands    & UTwente (1133)        & --   & 1      \\
\rowcolor{cyan!10}
us-los    & Los Angeles\newline United States & USC (4)             & --   & 1      \\
us-mia    & Miami\newline United States     & Ampath (20080)        & --   & 1      \\
\rowcolor{cyan!10}
us-was    & Washington\newline United States  & Los Nettos (226)    & --   & 1      \\ 
nl-arn    & Arnhem                          & SIDN (1140)           & --   & 1   
\\\cline{1-5}
  \end{tabularx}}
  \caption{Tangled sites location and connectivity.}
  \label{tab:tangled-nodes}
\end{table}

Since our goal is to create tailored experiments for anycast, we also
have implemented tools for controlling and measuring systems. These
tools are described in the next sections.

\section {Traffic Engineering on Tangled}
\label{sec:tagled-TE}
IP anycast relies on BGP for the routing of users' traffic to the
available anycast sites; in this context, the optimal situation is
typically defined as users being routed to the topologically nearest
anycast site. One of the challenges for anycast operators is not
having complete control  on catchment because of the complexity and
limitations of BGP routing. However, BGP does have mechanisms to
express routing preferences, ultimately influencing routing decision
processes. For example, one can prioritize some paths over others. In
\tangled, we support two methods of BGP engineering: AS-path
manipulation and community strings.

\textbf{AS-path manipulation} lies in making changes in the BGP path
attribute. AS-path attribute is used to implements loop avoidance in
BGP. An AS-path carries a list of all ASes from the current site, back
to the route originator, providing a rough distance estimation metric
measured in number of AS hops. The AS-path manipulation can be done
by: (1) \emph{prepending}, decreasing the preference of a routing path
by inflating its number of hops; (2) \emph{poisoning}, indicating ASes
to oppose a given path; or (3) \emph{reverse prepending}, by inflating
all but one paths.

\textbf{Community String} is a label optionally informed with the
prefix announcement, which is interpreted by the BGP neighbor and
translated into an internal AS routing policy. Communities are widely
supported by ISPs to delegate some of the BGP routing control to their
customers. Although community labels are not standardized, some
conventions do exist; for example, \emph{well-known communities} map
labels to routing policies such as \emph{black-holing} and
\emph{no-export}~\cite{rfc8642}. Communities can be propagated to all
the neighbors of a BGP router, or can target a particular AS.
\emph{Selective communities} are those that allow a specific routing
policy to be applied only to one individual selected AS.
\autoref{tab:tangled-moves} summarizes the BGP communities available
in \tangled.

We classify the available community strings in \tangled in the following 
routing policies:

\begin{itemize}
  \item \emph{Prepend}: send an inflated AS-Path to a neighbor
  \item \emph{noPeer}: do not send prefix information to IXPs or private peering
  \item \emph{noExport}: do not propagate this announcement beyond the neighboring AS
  \item \emph{noClient}: do not send this prefix to ISP customers
  \item \emph{Selective Prepend}: ask to upstream/IXP to prepend our prefix when sending to a specific AS neighbor
  \item \emph{Selective Advertise}: send prefix only to a specific AS; or, send to all but a specific AS
\end{itemize}

\autoref{tab:tangled-moves} shows that there is no homogeneity among
\tangled's transit providers in terms of BGP community coverage. Such
differences among ISPs is not considered an actual problem; it is
rather a reflection of the freedom that ISPs have on defining how to
support their respective clients.

\begin{table}
  \small
\scalebox{0.85}{%
\begin{tabularx}{\textwidth}{p{12mm}|p{3mm}p{3mm}p{3mm}p{3mm}p{3mm}p{3mm}p{3mm}p{3mm}p{3mm}p{3mm}p{3mm}p{3mm}}
\multirow{2}{10mm}{\textbf{Routing}\newline \textbf{Policy}} & \multicolumn{12}{c}{\textbf{Anycast Sites}}  \\
            & arn & cop & ens  & gru  & hnd & lnd & los   & mia  & par  & poa
            & syd & was \\ \cline{1-13}
\cline{1-13}
  \rowcolor{cyan!10} 
Prepend     & \ok & \ok & \ok  & \ok  & \ok & \ok  & \ok  & \ok  & \ok  & \ok  & \ok  & \ok \\ 
noPeer      & --  & --  &  --  & \ok  & --  & \ok  & --   & \ok  & \ok  & \ok  & \ok  & -- \\ 
  \rowcolor{cyan!10}
noExport    & --  & --  &  --  & \ok  & --  & \ok  & --   & \ok  & \ok  & \ok  & \ok  & -- \\ 
noClient    & --  & --  &  --  & \ok  & --  & --   & --   & \ok  & --   & --   & --   & -- \\ 
  \rowcolor{cyan!10}
Selective \newline 
Prepend     & --  & --  &  --  & \ok  & --  & \ok  & --   & \ok  & \ok  & \ok  & \ok  & -- \\ 
Selective \newline 
Advertise   & --  & --  &  --  & \ok  & --  & \ok  & --   & \ok  & \ok  & \ok  & \ok  & --  \\ 
\cline{1-13}
\end{tabularx} }%
  \caption{Traffic Engineering options on each site}
  \label{tab:tangled-moves}
\end{table}

\subsection{Inter-domain Routing Programming}

To simplify the routing management across the anycast sites in
\tangled, we developed an open-source tool named \emph{tangled-cli}.
Built on top of ExaBGP~\cite{thomas2010exabgp}, one can use
\emph{tangled-cli}'s interface to manage anycast site individually:

\begin{itemize}
    \item perform regular BGP prefix site announcements
    \item withdraw the BGP prefix from any site
    \item performing AS-path prepending
    \item announce a specific community string to a neighbor
    \item get the configuration of all active anycast sites
    \item get the status of all BGP peers
\end{itemize}

\autoref{lst:vpcli} shows examples of BGP routing configuration from
the \emph{tangled-cli}. The first command line configures a prefix
announcement using the IPv6 prefix 2001:610:9000::/40 from the anycast
site fr-par-anycast. In the second command line, we configure 20 path
prepending on the IPv4 prefix 145.100.118.0/23 for the anycast site
br-poa-anycast.

In addition to prepending and community strings, \emph{tangled-cli}
has other functionalities to help manage the anycast sites, such as
list prefix, remove BGP policy, and withdraw BGP prefix.

\begin{lstlisting}[language=bshell,fontadjust,
captionpos=b, caption={tangled-cli interface},label={lst:vpcli}]
$ tangled-cli -6 -A -t fr-par -r 2001:610:9000::/40
$ tangled-cli -4 -A -t br-poa -r 145.100.118.0/23 -P 20   
\end{lstlisting}

\section {Data Measurement and Analysis}
\label{sec:data}
There are multiple ways to measure anycast networks towards studies of
performance and behavior
\cite{RN276,ballani2006measurement,huang2008measuring,de2017anycast}.
In the case of \tangled, we deployed
\verfploeter~\cite{de2017anycast}, which we describe next.

\subsection{Anycast Mapping Measurements}

\begin{figure}
  \includegraphics[width=1.0\linewidth] {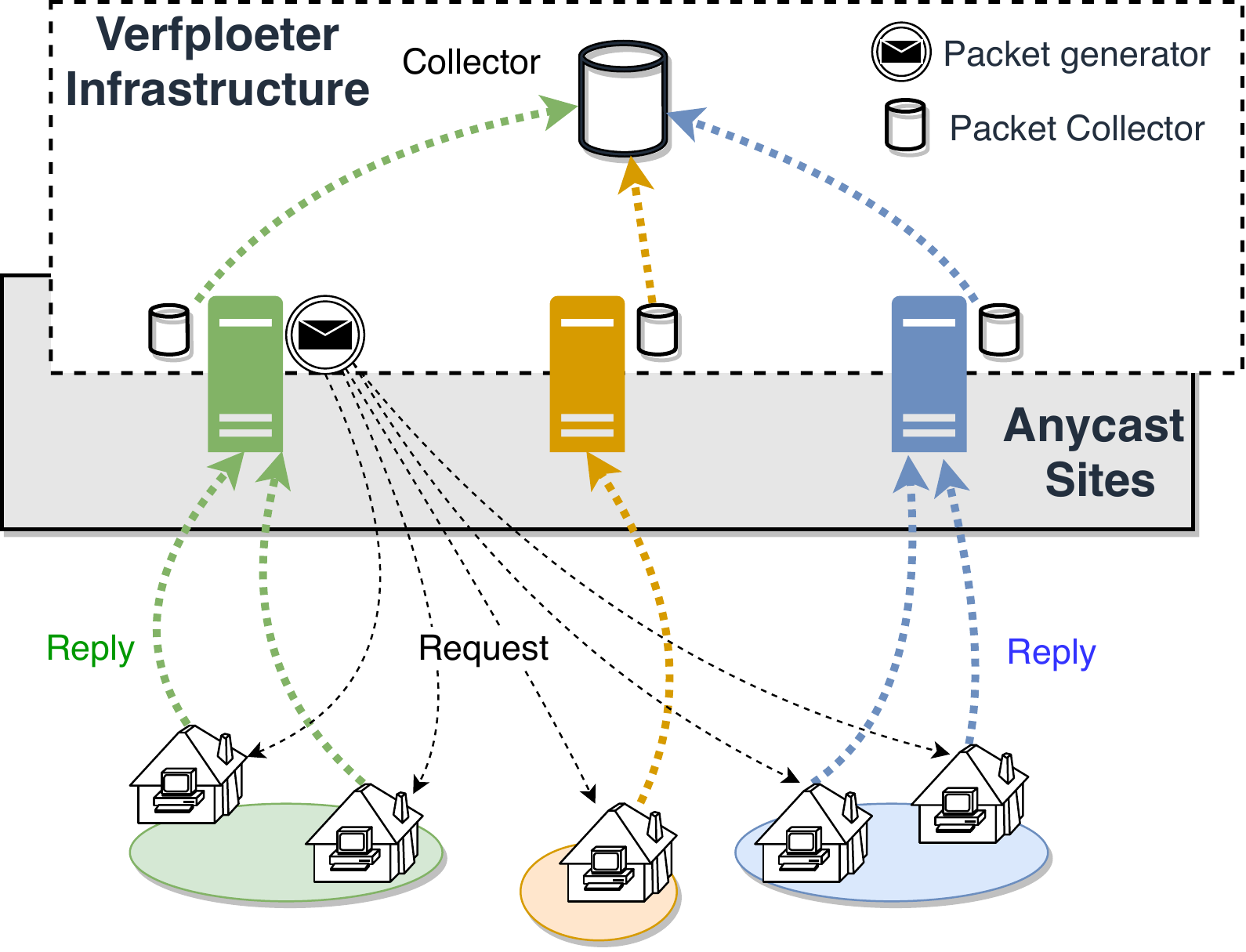}
    \caption{Verfploeter and its vantage points.} 
  \label{fig:tangled-verfploeter}
\end{figure}

\verfploeter\ actively probes IP addresses within a hit list (vantage
points--VP) using ICMP ECHO requests, to map clients of a distributed
service which is configured with IP anycast.
\autoref{fig:tangled-verfploeter} shows the catchment mapping
extracted from \verfploeter. ICMP ECHO requests are sent by one or
more servers called \emph{Pingers}; these servers may be, for example,
actual anycast sites or other multi-purpose servers.

The source IP address used in the ICMP ECHO messages is the address
configured in the anycast service. Active VPs replying to the ICMP
request, set the destination IP address of their respective ICMP REPLY
messages to that of the anycast service. Therefore, anycast sites will
receive ICMP REPLY messages without actually sending an ICMP ECHO
request. The set of received replies by each site defines their
respective anycast catchment.

\textbf{Measurement Duration.}

The duration of an entire measurement depends on how large is the IP
hit list, and also how frequent the ICMP ECHO requests are sent out to
their destinations as well as how many \emph{Pingers} are actively
probing. One could easily probe the entire set of valid /24 networks
within the Internet in minutes---our estimations is of 30 minutes for
a measurement with just one \emph{Pinger} and 6,5 millions IP
addresses in the hit list. However, we strongly take care of
measurements that send large amounts of ICMP requests within a short
period of time because they can be understood as an abusive behavior.
As described in~\cite{fan2010selecting}, actively probing hosts in the
Internet should not generate traffic that is discernible from the
\emph{traffic background noise}.

\textbf{Vantage Points.}

The accuracy of measurements in \verfploeter\ strongly depends on the
number and distribution of VPs, and also on how responsive they are.
Examples of hit lists that can be used in \verfploeter\ are those
built in~\cite{fan2010selecting}~\cite{b-dataset}, or an Alexa's
top-sites listing. In addition, geolocation of VPs can be based on any
geoIP database/source of choice.

\textbf{Catchment and Traffic Load.}

Since each VP in a hit list can be mapped to a /24 network, we can
estimate the traffic load that each anycast site would receive in an
actual operation. The accuracy of such an estimation, however, depends
on how comprehensive the VPs hit list is. Moreover, if unknown, the
distribution of traffic origins in such estimation would have to be
uniform across all /24 networks.

\textbf{Latency Measurements.}

To enable latency measurements, \verfploeter\ inserts a timestamp on
each outgoing ICMP ECHO request. When the ICMP REPLY is received at
one of the anycast sites, the difference between the first timestamp
and the receiving time is recorded. This time difference is a
triangular round-trip-time, similar to that of RTT concept.

The IP TTL information is also collected. Sample measurement data is
presented in \autoref{tab:verfploeter_data}.

\setlength{\tabcolsep}{6pt} %

\subsection{Data Analysis}

To explore the data generated from the anycast measurements, we have
developed tools to support that analyze. In particular, we are
interested on analyzing the data produced by \verfploeter aiming to
find the traffic distribution and catchment.

Each round of measurement probes more than 6 millions of networks and
generates around 400MB uncompressed text data.
\autoref{tab:verfploeter_data} shows a summarized view of the
measurement output. All data is exported in comma separated value
(CSV) format so to be easily interchanged.

\setlength{\tabcolsep}{5pt} %
\begin{table}[h!]
  \scalebox{1.0}{%
\begin{tabular}{crrrrrr}
    \textbf{Site}  & \textbf{Time Diff} & \textbf{Target IP}& \textbf{Anycast IP}& \textbf{TTL} & \textbf{CC}&  \textbf{ASN} \\ \hline
    au-syd & 97.191805    &  1.1.1.2          &  145.100.118.1     &   52         &    AU      &     13335     \\
    au-syd &102.285587&1.0.0.230&145.100.118.1&52&AU&13335 \\
    au-syd &110.469751&1.0.7.1&145.100.118.1&52&AU&56203 \\
    au-syd &116.260893&1.0.4.4&145.100.118.1&52&AU&56203 \\
\end{tabular}} %
  \caption{Anycast catchment measurement data provided by Verfploeter.}
\label{tab:verfploeter_data}
\end{table}

To help deal with such amount of data, we provide a tool to quickly
parse data provided by \verfploeter output and present the catchment
distribution. \autoref{lst:vpcli2} show an example. The listing shows
an anycast service using 6 sites and the respective number of replies
that each site handled during the measurement. The site
\textit{us-los-anycast-01} has received 1,342,542 replies, which
represent 37\% of queries performed in the measurement. This means,
that 37\% of clients reach the mentioned site.

%
%
\begin{lstlisting}[language=bshell,fontadjust,captionpos=b, caption={Quick
Tangled data analysis overview},label={lst:vpcli2},escapeinside={(*}{*)}]
# sites| replies -  percentual

us-los | 1342542 -  37%  (*\ding{110}\ding{110}\ding{110}\ding{110}\ding{110}\ding{110}\ding{110}\ding{110}\ding{110}\ding{110}\ding{110}\ding{110}\ding{110}\ding{110}\ding{110}\ding{110}\ding{110}\ding{110}\ding{110}\ding{110}\ding{110}*)
uk-lnd | 1123535 -  31%  (*\ding{110}\ding{110}\ding{110}\ding{110}\ding{110}\ding{110}\ding{110}\ding{110}\ding{110}\ding{110}\ding{110}\ding{110}\ding{110}\ding{110}\ding{110}\ding{110}\ding{110}*)
us-mia |  541846 -  15%  (*\ding{110}\ding{110}\ding{110}\ding{110}\ding{110}\ding{110}*)
fr-par |  473867 -  13%  (*\ding{110}\ding{110}\ding{110}\ding{110}\ding{110}*)
au-syd |   85475 -   2%  (*\ding{110}*)
jp-hnd |     321 -   0%  (*\ding{120}*)

\end{lstlisting}

A commonly used method to analyze data is using Jupyter notebooks
\footnote{\url{https://github.com/joaoceron/verfploeter-ttl-investigation}}
\footnote{\url{https://github.com/LMBertholdo/BQ-rtt}}.
  
\begin{figure}
  \includegraphics[width=1\linewidth] {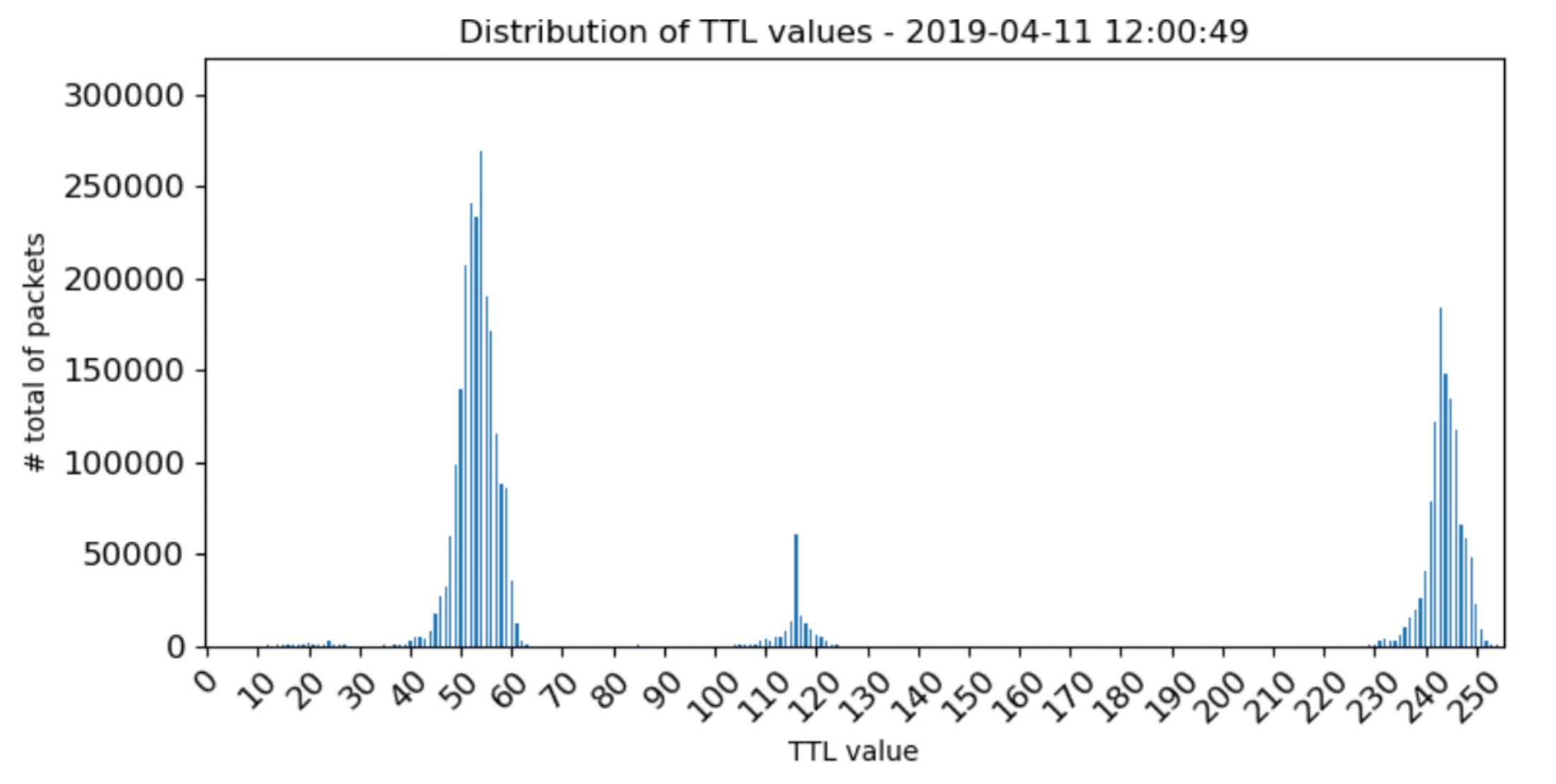}
    \caption{Collected TTL distribution on millions vantage points.} 
  \label{fig:ttl}
\end{figure}

\begin{figure}
  \includegraphics[width=1\linewidth] {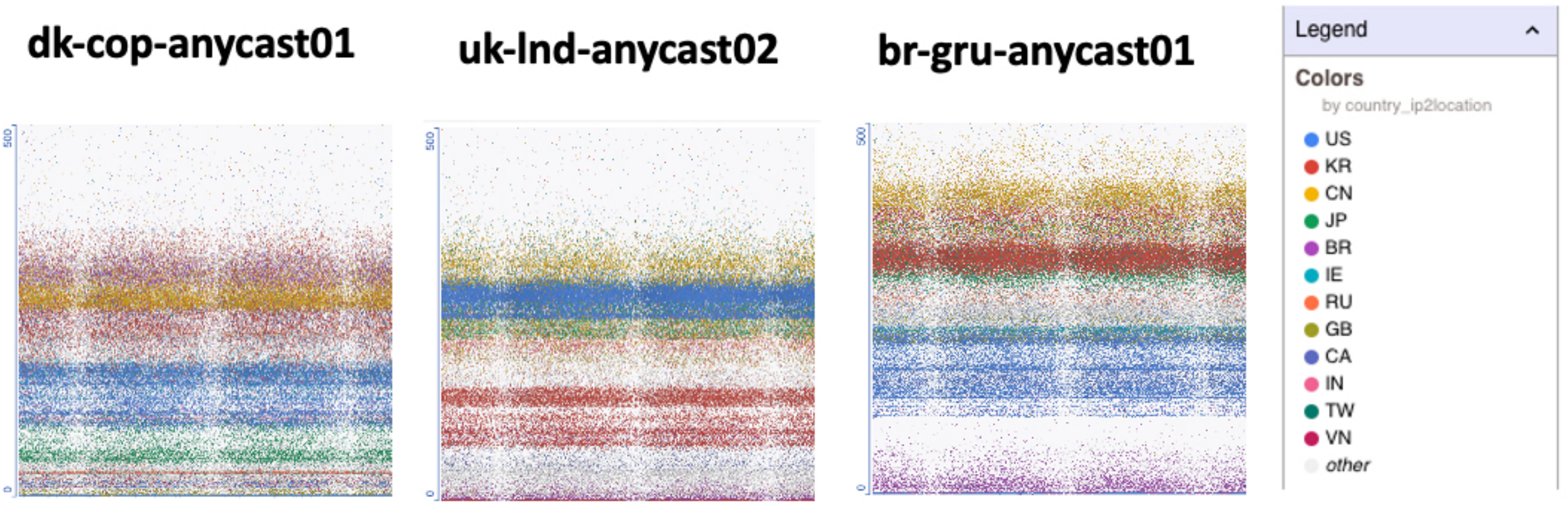}
    \caption{Round-trip-time by site and country generated using Google BigQuery.} 
  \label{fig:BQ-rtt}
\end{figure}

In \autoref{fig:ttl}, we inspect the time-to-live of all packets
received in one measurement round, totaling 4.5 million vantage points
answers. However, regular measurements can easily lead to big data
problems, demanding to analyze a huge amount of data. To support this
kind of investigation, we have written some codes able to upload
measurement and use big data solutions, such as the \emph{Google Big
Query} platform. One example of this is the round-trip-time analysis,
shown on  \autoref{fig:BQ-rtt}. This figure show individual
round-trip-time of million different vantage points, which site each
one are choosing, and in which country this vantage point is located.

\section{Lessons Learned}
\label{sec:lessons}

While running \tangled, we identified some challenges and learned some
lesson related to running a testbed as a service. First, the Internet
is dynamic and things breaks and get fixed without notification --
After a year of operation, we noticed that our providers changed
upstream and not properly announced our blocks to new provider, or our
peer made mistakes changing routing policies and affecting our
routing, or equipment replacement on our provider degrading our
overall performance. Our first lesson learned is that we need to
implemented a baseline checkup on all nodes first any measurement has
been taken as part of our management process. Second, inter-domain
routing has a slow convergence -- when we use to define our
inter-domain routing by software (SDN), the full time of BGP and
forwarding plane of all routers on Internet is slow, around 10
minutes. Third, collaborative testbeds as \tangled have some
drawbacks. Since most of our anycast sites are deployed and maintained
by hosting partners, in a best-effort fashion, we have observed some
limitations related to the operation of the infrastructure itself, as
well as to keep running long-term measurements. In general we register
issues related to:

\begin{itemize}
    \item lack of peering control: we are always submitted to our 
    collaborator routing policy.
    \item packet loss due to uncontrollable and unforeseen networking 
    issues as changes on quality of service policies of temporary 
    rate limiting.
    \item unpredictable (temporary) unavailability of anycast sites
    \item storage and bandwidth capacity are not unlimited on 
    local sites. Tests  involving high volume need to be evaluated 
    before (\aka\ DDoS studies).
\end{itemize}

Limitation we registered mostly affected long term measurements.
However, we have learned that carefully planning measurements
circumvent problems such as temporary unavailability of anycast sites.

\section{Related Work}
\label{sec:related}

Anycast research can be carried out by using simulators
\cite{katabi2000gia}\cite{agarwal2001content}, testbeds
\cite{sarat2006use}\cite{devries2017verfploeter}\cite{rizvi2020anycast}
or anycast networks in production
\cite{mcquistin2019taming}\cite{de2020global}. Anycast simulations are
used in specific cases when you need to study site load and swarm and
mobile catchment behaviors, usually in mobile and wireless networks
\cite{amir2020anycast}. Anycast testbeds are normally used to
Internet-related CDNs, DNS, and DDoS studies
\cite{devries2017verfploeter}\cite{rizvi2020anycast}.

Three distinct testbeds have been used for anycast tests so far. The
first one is \emph{Planetlab}~\cite{chun2003planetlab}, a testbed for
overlay networks used to develop \cite{freedman2006oasis} a global
anycast solution. Other is \emph{Peering}\cite{schlinker2014peering},
a BGP testbed widely used in Internet's BGP routing system research
and for some anycast research
\cite{dainotti2016hackatoon}\cite{li2018anycast}\cite{rizvi2020anycast}.
The last one is \tangled, a testbed specific for anycast research and
test. Over several anycast studies are carried out by
\cite{ripe73wouter}\cite{de2016anycast}\cite{kuipers17}\cite{de2020global}\cite{ceron2020bgptuner}\cite{sand2020ceron}\cite{paaddos2020bertholdo}\cite{rizvi2020anycast}.

Even though it is possible to built one's own testbed even by renting
capacity from some anycast or cloud provider; the whole anycast
measurement setup for data collection still has to be built. In
general the process of setting up, testing, and validating the whole
testbed environment spend months. Instead of wasting time building
one's own testbed, now researches can easily run their own anycast
experiments and focus on improving their ideas and results.

\section*{Acknowledgement}
\label{sec:ack}

This project have the support of SIDNLabs and NSNetLabs and is founded by DHS
  HSARPA Cyber Security Division via contract number
  HSHQDC-17-R-B0004-TTA.02-0006-I, Netherlands Organisation for scientific
  research NWO and CONCORDIA, the Cybersecurity Competence Network supported by
  the European Union\textquotesingle s Horizon 2020 research and innovation
  programme under grant agreement No 830927.

\bibliographystyle{bibliography/IEEEtran}
\bibliography{bibliography/ref.bib}

\end{document}